\documentclass[12pt,preprint]{aastex}

\def\dist{(m-M)_0}
\def\distv{(m-M)_V}

\newcommand{\mean}[1]{\langle #1 \rangle}

\shorttitle{Detection and Photometry of Clusters}
\shortauthors{Dolphin \& Kennicutt}

\begin{document}

\title{HST Survey of Clusters in Nearby Galaxies.\\
I.  Detection and Photometry}
\author{Andrew E. Dolphin}
\affil{Kitt Peak National Observatory, National Optical Astronomy Observatories, P.O. Box 26372, Tucson, AZ 85726}
\email{dolphin@noao.edu}

\and

\author{Robert C. Kennicutt, Jr.}
\affil{Steward Observatory, University of Arizona, Tucson, AZ 85721}
\email{robk@as.arizona.edu}

\begin{abstract}
We have developed photometric techniques that can be applied to images with highly-variable backgrounds, as well as to slightly-extended objects (object size comparable to or smaller than PSF size).  We have shown that ordinary stellar PSF-fitting photometry can be applied to slightly-extended objects, provided that one applies a systematic correction to the photometry that is a function primarily of the observed sharpness.  Applying these techniques to the Cepheid target NGC 3627, we find that we are successfully able to photometer the stars and clusters, as well as discriminating the cluster population with a negligible number of false detections.
\end{abstract}


\section{Introduction}
For decades it has been well established that the population of star clusters in the Milky Way is strongly bimodal, with a population of old, massive, and compact globular clusters in the Galactic spheroid and a physically distinct population of less populous open clusters in the disk.  However a growing body of observations suggests that this distinction is less pronounced in other types of galaxies, and that the nature of the young and intermediate-age cluster populations differs dramatically in different galactic environments. Early studies of the star cluster populations in the Magellanic Clouds revealed the presence of young and intermediate age ``populous blue clusters'' or ``blue globular clusters'' that are absent in the Milky Way \citep{hod61}, and similar populations have been found in M33 and other nearby late-type galaxies \citep{chr82, ken88a}.  Populations of very massive ``super star clusters'' have been found in starburst galaxies and strongly interacting galaxies \citep{hol92, ocon94, sch96, whi99, zep99}, leading to the speculation that many of these objects are progenitors of globular clusters \citep{ash92}.

Other observations obtained over the past 15 years show that the specific frequencies and mass spectra of young stellar associations vary systematically along the Hubble sequence.  The most complete data of this kind come from the luminosity functions of giant HII regions, which roughly trace (for luminous objects) the ultraviolet luminosity functions of the embedded stellar associations.  The luminosities of the brightest HII regions increase sharply from S0/Sa -- Sc and Irr galaxies \citep{ken88b,cal91}, due to the combination of an increase in both the total numbers of star forming regions per unit galaxy mass, and a change in the shape of the association luminosity function \citep{ken89,cal91,bre97,you99}.  Reliable measurements of the continuum luminosity functions of the associations themselves are much more difficult to measure, due to the increased difficulty of isolating the associations against the background light of the disk, but studies by \citet{bre97} and \citet{lar00} are consistent with the results derived from HII region statistics, namely that the frequency and characteristic luminosities of the associations increase in systems with higher star formation rates (SFRs).

Taken together these studies show that the clustering spectrum of star formation is sensitive to the local physical conditions in the interstellar medium.  Such variations are of critical importance for understanding galaxy formation and evolution, because the feedback effects of star formation on the ISM via radiation, supernovae, and stellar winds, can vary dramatically depending how concentrated the massive star formation is in space and time \citep{mac88, hei90, ike91, wil97}.

A critical missing link in this picture is a systematic study of how the populations of massive, bound star clusters vary as functions of galaxy type, ISM properties, or star formation rates.  Although the existing observations of selected Local Group galaxies, starburst galaxies, and merger remnants show strong differences in the properties of their young massive cluster populations, the data are fragmentary, and fundamental questions remain about the physical nature and origin of these differences.  For example it is not at all clear whether the ``super star clusters'' or ``blue globular clusters'' in starburst galaxies are the result of a distinct mode of cluster formation (analogous to the differences between the open and globular cluster systems in the Milky Way), or whether they simply are the extension of a continuous trend in disk cluster populations along the Hubble sequence.  Likewise it is unclear whether the strong trends seen in the properties of OB/HII associations in galaxies are also present in the bound cluster populations.  

The absence of a substantial body of data on clusters in nearby galaxies is not surprising, because a typical object ($M \ge 10^4 M_\odot$, $R < 10$ pc) becomes effectively stellar in appearance beyond a distance of $\sim$1 Mpc when observed from the ground, and reliable identification and measurement of such clusters is challenging even with the Hubble Space Telescope (HST).  However HST does offer the capability to obtain reliable statistics on at least the upper end of the luminosity function of young clusters out to distances of $\sim$20 Mpc, if high S/N images are analyzed with precision PSF-fitting techniques, and extensive artificial star and cluster simulations are used to quantify the effects of spatial resolution, structured background, and stellar crowding on the cluster identification and photometry.

We have undertaken such a survey of the massive young star cluster populations in nearby spiral and irregular galaxies, using archival imaging from HST.  Although we are using a familiar tool (PSF-fitting stellar photometry) for this project, we are devoting this first paper of the series to our photometry techniques because of this application to slightly-extended objects.  Although our methods are optimized for the problem of detecting compact star clusters against a background of resolved stars, many aspects of the technique should be applicable to other problems where semi-resolved objects need to be detected against a complex background of point sources.  Section 2 is used to outline the theory behind the technique, and Section 3 shows a sample reduction of NGC 3627 (M66).  Paper II will describe the statistical analysis we have developed to analyze the color-magnitude-radius information.

Subsequent papers in this series will apply these techniques to a sample of 30 nearby ($d < 25$ Mpc) spiral and irregular galaxies, in order to quantify the trends in cluster properties along the Hubble sequence, and to a more distant sample of starburst and other peculiar galaxies, to better quantify any differences in the cluster populations in these objects.

\section{Analysis Technique}

Even with the superb resolution possible with the Hubble Space Telescope, the accurate detection and analysis of distant star clusters is a difficult process.  At 20 Mpc, a cluster with a core radius of 2 pc has a size smaller than the WFPC2 PSF.  After convolving the cluster and a stellar point source with the WFPC2 PSF, the cluster appears less than 20\% broader.  This difficulty in detection of clusters is further compounded by the two factors.  First, the relatively large pixels on WFPC2 (especially the wide field cameras) means that both the star and the cluster have PSFs smaller than one pixel; thus measurement of the FWHM of an object is thus impossible and size can only be estimated by the fraction of light falling outside the central pixel.  Second, blends of supergiants and distant background galaxies will also appear slightly broader than stars, and thus additional care must be taken when identifying cluster candidates.

\subsection{Photometry Algorithm}
\label{sec_phottype}

A very basic question that must first be addressed is the photometry method that we will employ.  There are three basic options: aperture photometry, stellar PSF-fitting photometry, and cluster PSF-fitting photometry.  Aperture photometry is the simplest; it involves the measurement of the number of counts within a radius of the center of the aperture.  As such, it is the least sensitive of the three to systematic errors, as no \textit{a priori} assumptions are made regarding size or shape of the object being measured.  However, aperture photometry is the most sensitive to variable background (as PSF-fitting photometry of any kind gives the highest weight to the central pixels), and most routines do not provide any measurement of star size or shape.  Thus aperture photometry is not the best algorithm for use in this project.

The other two possibilities are PSF-fitting routines of two flavors -- stellar and cluster.  Stellar PSF-fitting routines attempt a simple three-parameter fit for each object; the three parameters being position ($x$, $y$) and number of counts.  In addition to measuring these values for each object, sharpness and roundness values are also returned, generally for the purpose of rejecting non-stellar objects from the photometry list.  When searching for clusters, however, these values can also be used to reject non-clusters -- stars, blends, background galaxies, etc.  In contrast, cluster PSF-fitting photometry attempts a more complex fit of each object.  The routine described by \citet{lar99}, for example, fits six parameters per object: position, counts, and three size parameters related to major axis, minor axis, and orientation.

While the cluster-fitting approach has the advantage that it directly provides the size and shape of each object, the same information is also contained in the PSF-fitting algorithm output.  The only significant differences between the two approaches are the following.  Stellar PSF-fitting photometry has the disadvantage that there is a systematic error in the magnitudes caused by assuming a stellar PSF.  While this can be easily corrected, it does add an extra step in the reduction process.  Cluster PSF-fitting photometry, however, has the disadvantage that the full uncertainties are extremely difficult to measure -- generally only the standard error of the brightness measurement is reported.  Again, this problem can be solved by adding an extra step to the reduction process.

While we conclude that aperture photometry is not optimal for this project because of its large random errors, we find that either stellar PSF-fitting photometry or cluster PSF-fitting photometry will work equally well.  Because of our familiarity with the stellar PSF-fitting package HSTphot \citep{dol00a}, we will take that route for our primary reduction procedure, along with an updated version of its CTE corrections \citep{dol00b} available at AED's web site.  The process of photometering the clusters will thus be a two-step process -- stellar photometry with HSTphot, followed by determination of the systematic errors caused by use of the stellar PSF.

A final note involves the determination of the background.  The typical procedure is to measure the image values in an annulus around the object being measured and use a mode or robust mean algorithm to determine the background value.  However, in the galaxies in this survey, we face the difficulty of a rapidly-varying background in which the default HSTphot background annulus is much too large to provide an accurate measurement of the background value at the position in question.  Instead, we require a very crude technique in which the background is calculated by averaging the image values in a 2-pixel wide annulus immediately outside the photometry annulus.  This technique also avoids the bias inherent in most sky algorithms, which are based on the assumption that the lower values are more indicative of the true sky level, while higher values are caused by stars (which is generally true in uncroweded fields).  This also largely eliminates the effect of associations, as any association with a diameter of 5 WFC pixels or more (20 pc at a distance of 10 Mpc) will be subtracted out as background and any cluster within the association photometered properly.

\subsection{Stellar PSF Determination}
\label{sec_psf}

Regardless of whether one uses stellar or cluster PSF-fitting photometry, knowledge of the point source (stellar) PSF is the critical first step, as the observed image of any cluster is its intrinsic image convolved with the PSF.  For the data we are analyzing in this project, the determination of the stellar PSF is not a trivial process and the process should be described in detail.

HSTphot's base PSF library is computed using Tiny Tim PSFs \citep{kri93}.  The exact algorithm can be found in \citet{dol00a}; the end result is a library of many PSFs per chip per filter.  For the planetary camera, a total of 1600 PSFs per filter are measured -- $5 \times 5$ subpixel centerings at $8 \times 8$ chip positions.  Each wide field camera has a factor of four more PSFs, as there are $10 \times 10$ subpixel centerings at each chip position to compensate for the lager pixels (and thus larger differences between PSFs separated by a fraction of a pixel).

After reduction of many data sets, it was seen that the PSFs showed small systematic errors as a function of position on each chip.  This is assumed to be an error on the part of the Tiny Tim models, which while excellent, are obviously not expected to be perfect.  In order to compensate, each PSF taken from the library is expanded or contracted slightly (a scale difference of order 2 \%, independent of chip), using a quadratic correction term in addition to the expansion necessary to compensate for geometric distortion.  (Tiny Tim does not account for geometric distortion.)  The 34th row error is also corrected here, according to the prescription given by \citet{dol00a}.  These factors do not significantly affect our photometry (at the $\sim 2$\% level), but will significantly affect our cluster discrimination.

The result of this procedure is our best initial guess of the stellar PSF at any position on any chip.  However, small focus changes and small pointing errors will modify the stellar PSFs further, and it is necessary to add a correction image to the PSF.  This correction image is usually measured from bright, isolated stars, but such objects do not exist in many of our galaxy fields.  Instead, we frequently find that the only objects sufficiently bright for PSF determination are clusters; use of these to set the stellar PSF would clearly be disastrous if we are attempting to locate clusters by their size.  Instead, we adopt a two-iteration process in our photometry.

In the first iteration, we simply use the library PSFs, plus mean correction images determined from well-resolved Local Group dwarf galaxies, to photometer all stars.  As HST pointing errors (when returning to the same position on the sky) are generally much more significant than focus changes, this provides a good first-order approximation.  We then plot sharpness vs. magnitude for all recovered objects.  The locus of stars is usually obvious in such a plot; and a revised correction image can be calculated such that the mean star sharpness is zero.  (The HSTphot sharpness parameter is defined such that a clean star should have a sharpness of zero.)  The adjustment to the correction image is modeled as a difference between two PSF functions; one corresponding to the mean FWHM of stars in the image and the other a free parameter.  The functional form of the PSF used for this fit was
\begin{equation}
y = \frac{1}{(1 + c r^2)^{1.5}}
\end{equation}
for the PC and
\begin{equation}
y = \frac{1}{(1 + c r^2)^{2}}
\end{equation}
for the WFC; these functions provided the best fits to the PSFs.  In both equations, $c$ is the parameter being fit and $r$ is the distance from the center of the PSF.  Thus this is essentially a one-parameter fit that can be made for any galaxy in which even a small number of individual stars are resolved.  The photometry is then re-measured, using the adjusted PSF correction image, producing our final stellar photometry for the galaxy.

\subsection{Systematic Error Correction \label{sec_error}}

As mentioned in Section \ref{sec_phottype}, the principal weakness of stellar PSF-fitting photometry in this application is that the attempt to fit a slightly extended object using a stellar PSF will cause a systematic underestimation of the brightness.  This effect can be measured (and corrected) in two ways.  In this section we discuss an analytical approach; the next section discusses a semi-empirical approach.

Given the HSTphot equations given by \citet{dol00a}, we can calculate the expected photometric consequences of an object being slightly extended.  HSTphot uses \citet{dol00a} equation 8, reproduced here in equation \ref{eq_s}, to measure the signal contained in an object.  To eliminate gain from the equations, all values will be given in ADU.
\begin{equation}
\label{eq_s}
s = \frac{\sum_{x,y} wt_{x,y} R_{x,y} SPSF_{x,y} / \sigma_{x,y}^2}{\sum_{x,y} wt_{x,y} SPSF_{x,y}^2 / \sigma_{x,y}^2},
\end{equation}
where $wt_{x,y}$ is the weighting factor (generally 1; between 0 and 1 in the outermost pixels to set a circular aperture), $R_{x,y}$ is the residual image (original data minus sky and other nearby stars), $SPSF_{x,y}$ is the stellar PSF.  The expected scatter in a pixel, $\sigma_{x,y}$, is given by \citet{dol00a} equation 10,
\begin{equation}
\sigma_{x,y}^2 = R_{x,y} + sky_{x,y} + RN^2,
\end{equation}
where $sky_{x,y}$ is the mean sky level and $RN$ is the read noise in electrons.

For any object, the expected residual in any pixel is given by
\begin{equation}
R_{x,y} = s_0 PSF_{x,y},
\end{equation}
where $s_0$ is the actual number of counts contained while $PSF_{x,y}$ is the object's PSF, a convolution of the stellar PSF and the object's intrinsic shape.  Substituting this into equation \ref{eq_s}, we obtain
\begin{equation}
\label{eq_sratio}
\frac{s_0}{s} = \frac{\sum_{x,y} \frac{wt_{x,y} SPSF_{x,y}^2}{s_0 PSF_{x,y} + sky_{x,y} + RN^2}}{\sum_{x,y} \frac{wt_{x,y} PSF_{x,y} SPSF_{x,y}}{s_0 PSF_{x,y} + sky_{x,y} + RN^2}}.
\end{equation}
The magnitude correction is simply $dm = -2.5 \log(s_0/s)$, so the systematic error (as a function of actual PSF, brightness, and background) can be calculated using equation \ref{eq_sratio}.  Likewise, the measured sharpness (and indication of object size) is given by \citet{dol00a} equations 14 and 15, and is a function of actual PSF, brightness, and background that can be easily evaluated.

Because $R_{x,y}$ is randomly distributed with an uncertainty of $\sigma_{x,y}$, the magnitude error and sharpness measured this way are expectation values only.  The full distribution can most easily be measured using a Monte Carlo test, as $R_{x,y}$ appears in both the numerator and the denominator of equation \ref{eq_sratio}.

\subsection{Artificial Clusters}

The more direct approach, in terms of simplicity, is to place a large number of artificial clusters on the image and photometer them -- a procedure analogous to artificial star tests in stellar population work.  By placing many artificial clusters with a given magnitude and size on each frame, one finds not only the systematic difference between input and output magnitudes as a function of input parameters, but also the completeness fraction, photometry scatter, and mean sharpness.

We have developed a very detailed artificial cluster algorithm.  We began by attempting to reconstruct the young LMC cluster NGC 1818, using \citet{gir00} isochrones, a binary fraction of 35\%, the WFPC2 PSF, and the profile measured by \citet{els87},
\begin{equation}
L(r) = L(0) ( 1 + \frac{r}{9}^2 )^{-1.2},
\end{equation}
where $r$ is the distance from the cluster center in arcsec and $L(0)$ is the central surface brightness.  Using these inputs, we were able to quantitatively reproduce the observed CMD and spatial distribution of stars seen in archival WFPC2 images (proposal GO-6277).  Confident in this approach, we applied the same technique to add artificial clusters to our NGC 3627 images.

To simplify the problem (and because clusters at the distances we are observing are too small to determine the shape of the profile), we reduced the two structural parameters of \citet{els87} equation 1 to one parameter by noting that most of the scatter in their measured $\gamma$ values (their table 1) is from measurement uncertainties; the values are consistent at the $1.1 \sigma$ level with $\gamma = 2.53 \pm 0.10$.  Thus we will assume a universal cluster profile of
\begin{equation}
L(r) = L(0) ( 1 + \frac{r}{a}^2 )^{-1.26}.
\end{equation}
Converting to core radii ($r_c$) using \citet{els87} equation 22, this becomes
\begin{equation}
L(r) = L(0) ( 1 + 0.73 \frac{r}{r_c}^2 )^{-1.26}
\end{equation}
and their equation 2 becomes
\begin{equation}
L(<r) = L \lbrack 1 - ( 1 + 0.73 \frac{r}{r_c}^2 )^{-0.26} \rbrack,
\end{equation}
where $L(<r)$ is the luminosity falling within a radius and $L$ is the integrated luminosity.

The artificial cluster tests are thus made by placing clusters with a variety of ages, initial masses, and core radii on the image and photometering them.  We tests $\sim 20,000$ such clusters in our NGC 3627 reduction, described below.

We will use these two techniques in very different ways, as each has its strengths and weaknesses.  The artificial cluster approach is less prone to error, as one is actually photometering objects on the image.  By building up a large library of artificial cluster results, therefore, one can reproduce the observational effects with great accuracy.  This will be critical in our statistical analysis, the topic of Paper II.  However, orders of magnitude more artificial clusters are required in order to determine the integrated magnitudes and core radius of an observed cluster, a task that is done much more easily through our analytical technique.  Thus the analytical technique is used to convert observed properties to physical properties; the artificial cluster technique to do the reverse.  A comparison will be made in our NGC 3627 analysis.

\subsection{Identification and Measurement of Cluster Candidates}

Armed with HSTphot photometry and an understanding of how extended PSFs affect the photometry, we now attempt to determine which objects in our photometry list are likely clusters.  A library of artificial star tests (placed on the image with a distribution based on the distribution of light on the images in order to correctly sample the observed data) will help to identify the distribution of single stars and blended stars on a sharpness vs. magnitude plot; we can then identify outlying objects on the broad (negative sharpness) side of the diagram as clusters.  As we are primarily interested in finding compact clusters (which are round), we can also use roundness information, when available, to help reduce the number of spurious detections caused by blended stars and background galaxies.

Each candidate cluster is characterized by photometry (typically $VI$), standard errors in the magnitudes, sky brightness in each frame, and a sharpness measurement.  As sharpness is primarily a function of angular size, this value can be determined with reasonable accuracy, which in turn allows a measurement of the integrated $VI$ magnitudes (as described in the previous section).

In terms of studying the properties of the cluster population, however, it is more accurate to statistically model the observed data using theoretical models of the cluster population and observational errors as determined by ``artificial cluster'' tests.  The reason is one of accuracy; it is far simpler to use the artificial cluster results (completeness, systematic errors, and scatter) to create a simulated luminosity function and/or color distribution from a model population than it is to deconvolve these corrections from observed data.  This approach is identical in concept to the CMD-synthesis approach for studying stellar populations described by \citet{dol01b}, and will be explored in Paper II.

We will employ both types of corrections in this study.  In general, the properties of a single cluster are best described by the former; while statistical properties can be most easily described by the latter.

\section{Sample Reduction -- NGC 3627}

To illustrate our reduction and analysis procedures, we present a detailed description of our study of NGC 3627, using data studied previously by \citet{sah99} and \citet{gib00}.  One of the reasons for this selection is that \citet{gib00} found unusual aperture corrections when studying this galaxy, one possible explanation for which was a large cluster population.  Thus, while we expect to examine a large number of objects with no \textit{a priori} knowledge of cluster properties, NGC 3627 seemed an intriguing prospect for a first test of our cluster-detection methods.

WFPC2 observations of NGC 3627 were taken over a 60-day period between 12 November 1997 and 10 January 1998 for proposal GO-6549 (PI Sandage).  The data consist of 24 exposures in F555W (totaling 58800s of integration) and 10 exposures in F814W (totaling 25000s of integration), all images taken in pairs to allow the creation of 17 cosmic ray cleaned (12 F555W and 5 F814W) single-epoch images.  Unfortunately (for the purposes of creating a deep image), the images were split between two pointings separated by a few WFC pixels.  We therefore created two deep images in each filter using the \textit{crclean} routine of HSTphot \citep{dol00a}, in addition to the 17 single-epoch frames.  One of the deep F555W images is shown in Figure \ref{fig_image}.
\placefigure{fig_image}

\subsection{PSF Determination}

Because of the lack of bright, isolated stars in the data, we have used the iterative procedure described in Section \ref{sec_psf} to determine the stellar PSF.  An initial photometric pass was made using the \textit{multiphot} routine from HSTphot, producing stellar photometry (F555W and F814W magnitudes and sharpness) from the default PSFs.  The default options for HSTphot were used, except that the local sky value was calculated before each photometry measurement to compensate for the variable background, object classification was turned off to produce more uniform photometry, and aperture corrections and PSF corrections were turned off because of a lack of adequate stars.  The ``local sky'' algorithm is described in section 2; we are confident of its accuracy because we are able to recover artificial stars to a systematic accuracy of better than 1\% (systematic) down to $V = 25$ in WFC2 and $V = 26$ in the other chips.

Figure \ref{fig_psf0} shows all PC detections from one of the deep F555W images, with sharpness on the $y$-axis and $V$ on the $x$-axis.  (We choose $V$ because the amount of exposure time in $F555W$ was 2.4 times that in $F814W$.)  For comparison, Figure \ref{fig_psfngc121} shows a similar plot for an image centered on the SMC globular cluster NGC 121, which has more severe crowding but no clusters.  It is clear from a comparison of the two figures that a large population of extended objects exists in NGC 3627, which are distinguished by the negative sharpness values.  That this is not simply the result of a very broad stellar PSF in these data is demonstrated by the sharpness values of Cepheids found by \citet{sah99} and \citet{gib00} (assuming that most of the objects they classified as Cepheids are single stars), which are marked with asterisks in Figure \ref{fig_psf0} and largely fall near a median sharpness $-0.10$ (compared with a median sharpness of $-0.22$ for all objects in the PC).  Similar plots for the other three chips, as well as those for the other images, are qualitatively identical, with a stellar population similar to that in Figure \ref{fig_psfngc121} plus a population of extended objects.  Eliminating the Cepheids that are outliers and using the remainder as PSF stars, we measure the mean values of sharpness, background, and counts, which are given in Table \ref{tab_psf0}.  From these values we calculate the PSF adjustments necessary to produce zero mean sharpness, given as the final column of Table \ref{tab_psf0}.  Comparing the corrections for the two pointings (assuming the actual PSFs are the same in each pointing), we estimate the rms error in the FWHM corrections to be $\pm 0.09$ pixels.
\placefigure{fig_psf0}
\placefigure{fig_psfngc121}
\placetable{tab_psf0}

After determining an appropriate PSF adjustment, the photometry was redone.  Figure \ref{fig_psff} shows the revised plot of sharpness values, with the crosses again indicating our PSF stars (the Cepheids found by previous studies).  Although there is some scatter, the PSF stars have a median sharpness of $-0.05$, significantly improved from the first iteration.  (That the sharpness is not exactly zero after applying the corrections from \ref{sec_error} is a result of the interdependence of the background level measurement and PSF in the photometry algorithm.  An additional iteration would reduce the median sharpness further, but $-0.05$ is adequate for our purposes.)
\placefigure{fig_psff}

\subsection{Distance and Extinction}

One of the selection criteria mentioned in the introduction is that we require the deepest possible images so that cluster sizes can be measured with reasonable signal-to-noise.  A majority of the data sets meeting this criterion were taken in Cepheid searches, which require of order 30 deep images in one filter.  Because of this, there are two by-products of this cluster survey.  First is the opportunity to make a consistency check on our stellar photometry by comparison with that of the previous groups to have analyzed these data.  Second is that these objects have well-measured distances and mean extinctions as a result of the Cepheid searches, which will thus improve the accuracy of our cluster measurements.

Thus, although we are searching for clusters, a by-product of our NGC 3627 stellar photometry is photometry of the Cepheids found by \citet{sah99} and \citet{gib00}.  We obtained light curves for the previously-discovered Cepheids using techniques described by \citet{dol01}.  Mean magnitudes and periods are given in Table \ref{tab_cepheid}.  For the six Cepheids well-recovered in all three studies (using a cutoff of $\hbox{Q} \ge 3$ for our photometry), we find excellent agreement in the reddening-free magnitudes ($W = 2.45 I - 1.45 V$) -- the present photometry and that of \citet{sah99} agree to $0.00 \pm 0.04$ magnitudes; \citet{gib00} is $0.02 \pm 0.06$ magnitudes fainter.  We note that our magnitudes are systematically fainter than the other studies ($\sim 0.2$ magnitudes in $V$ and $\sim 0.1$ in $I$); the cause of this is likely our technique for determining sky values over a rapidly-varying background.
\placetable{tab_cepheid}

We can determine the distance and extinction of NGC 3627 using the $V$ and $W$ magnitudes and a Cepheid period-luminosity (P-L) relation.  Adopting the \citet{mad91} P-L relations, we find a true distance modulus of $\dist = 30.09 \pm 0.10$ and an apparent distance modulus of $\distv = 30.67 \pm 0.07$ (and thus a mean extinction of $A_V = 0.58 \pm 0.12$).  This is consistent with the literature values of $\dist = 30.17 \pm 0.12$ \citep{sah99} and $\dist = 29.99 \pm 0.17$ \citep{gib00}; both literature values have been corrected to the zero points used in the present work.  Adopting the \citet{uda99} P-L relations (adjusted to an assumed LMC distance modulus of 18.50), we find a true distance modulus of $\dist = 29.97 \pm 0.09$ and an apparent distance modulus of $\distv = 30.73 \pm 0.07$ (a mean extinction of $A_V = 0.76 \pm 0.11$).  We find this distance inconsistent with the value of $\dist = 29.71 \pm 0.08$ reported by \citet{fre01}; the reason for the discrepancy is an unexplained subtraction of 0.18 mags from the \citet{gib00} distance.  We believe this is because of the aperture correction discrepancy mentioned by \citet{gib00}; if this is indeed the cause, the shorter distance of \citet{fre01} is in error.  For either relation, the scatter in $\distv$ is mostly accounted for by observational errors, implying that effects of differential reddening are present but not large.  We conservatively adopt the average values and increase the uncertainties accordingly, producing a distance modulus of $\dist = 30.03 \pm 0.11$ (corresponding to a distance of $10.1 \pm 0.5$ Mpc) and mean extinction of $A_V = 0.67 \pm 0.15$.

\subsection{Cluster Discrimination}

We finally turn our attention to the problem of locating the clusters.  We will again use the sharpness vs. $V$ plots to make this discrimination, beginning with the artificial star tests created by HSTphot's \textit{multiphot} routine.  From these tests, we find the $1 \sigma$ lower limit of the artificial star population to be given by
\begin{equation}
\hbox{sharpness} = -0.01 - 0.293 \times e^{0.46 ( V - 28 )}
\end{equation}
for PC1, WFC3, and WFC4; and
\begin{equation}
\hbox{sharpness} = -0.01 - 0.731 \times e^{0.46 ( V - 28 )}
\end{equation}
for WFC2.  Similar limits can be given for the HSTphot roundness parameter, which is zero for a perfectly-round object and 1 or higher for an elongated object.  The $1 \sigma$ limits for the artificial star populations are given by
\begin{equation}
\hbox{roundness} = 0.01 + 2.56 \times e^{0.46 ( V - 28 )}
\end{equation}
for PC1;
\begin{equation}
\hbox{roundness} = 0.01 + 4.44 \times e^{0.46 ( V - 28 )}
\end{equation}
for WFC2; and
\begin{equation}
\hbox{roundness} = 0.01 + 2.18 \times e^{0.46 ( V - 28 )}
\end{equation}
for WFC3 and WFC4.  Figure \ref{fig_psffake_envelope} shows the $2.5 \sigma$ sharpness envelopes of artificial stars in four chips; Figure \ref{fig_psffake_round} shows the $2 \sigma$ roundness envelopes.  An examination of Figure \ref{fig_psffake_envelope} uncovers two notable features.  First, that the plot is largely (though not completely) symmetric between positive and negative sharpness values indicates that blending of individual stars (thus creating the appearance of clusters) is not a major factor in these data.  We can quantify the contribution of single star blends and photometry errors to our cluster sample by measuring the ratio of the number of artificial stars recovered below the sharpness envelope but inside the roundness envelope to the number recovered within the both envelopes.  Limiting the photometry to $V \leq 23.5$ in WFC2 and $V \leq 24.5$ in the other chips to reduce contamination, we find contamination fractions of 0.14\% in PC1, 0.18\% in WFC2, 0.17\% in WFC3, and 0.23\% in WFC4.
\placefigure{fig_psffake_envelope}
\placefigure{fig_psffake_round}

A comparison of Figures \ref{fig_psfngc121} and \ref{fig_psffake_envelope} indicates that, for real data, we should add a constant width of 0.05 to the envelope.  (Observed data are unfortunately never as well-behaved as simulated data.)  This is done for our observed NGC 3627 data, which is plotted in full in Figures \ref{fig_psf_envelope} and \ref{fig_psf_round}.  In total, we have 923 ``stars'' (objects with sharpness and roundness values within the stellar envelopes), which, based on our contamination fractions from the previous paragraph, produce an estimated 1.7 false detections of clusters.  In contrast, we have 553 objects falling below the sharpness envelope but within the roundness envelope; the contamination is thus negligible.  It is clear that there is also a large population of extended objects at fainter magnitudes, although discrimination is impossible with the accuracy that we have obtained for brighter objects and thus we have chosen to omit them from our sample.
\placefigure{fig_psf_envelope}
\placefigure{fig_psf_round}

\subsection{Cluster Properties}

As described in section \ref{sec_error}, we have two methods of measuring cluster sizes and integrated magnitudes from HSTphot photometry -- theoretically or semi-empirically.  The theoretical approach attempts to convert observational quantities to physical properties; the artificial cluster approach is the opposite.  It is important that we test the consistency of these two techniques.

In order to do so, we have added nearly 20,000 artificial clusters to our NGC 3627 images, of which roughly a quarter were recovered with $V$ magnitudes and sharpness values that met our selection criteria for clusters.  (The low recovery rate is because our input clusters included many that were saturated, too faint to be detected, or too small to be classified as clusters.)  We show the sharpness and roundness values for artificial clusters with core radii of 2.5 pc in Figures \ref{fig_psffclus_envelope} and \ref{fig_psffclus_round}.  The relatively flat and sharp trends in Figure \ref{fig_psffclus_envelope} demonstrate that the HSTphot sharpness value is principally a function of cluster radius.  The completeness, as a function of both input magnitude and core radius, is shown in Figure \ref{fig_complete_pc}.  We note that there are limited ranges of both cluster radii and magnitudes to which we are sensitive.  The radii are limited by the 1-pixel limit on the large end (to minimize false detections of associations), and by the sharpness limit (a function of magnitude) on the small end.  The brightness range is restricted by our magnitude cut on one end and pixel saturation on the other.
\placefigure{fig_psffclus_envelope}
\placefigure{fig_psffclus_round}
\placefigure{fig_complete_pc}

The $V$ corrections and core radii from our analytical corrections, as functions of measured $V$ magnitude and sharpness, are shown (for typical PC background levels) in Figures \ref{fig_sampledv} and \ref{fig_samplesize}, respectively.  The key feature of both of this figures is that both the magnitude correction and core radius are primarily functions of sharpness, allowing a relatively straightforward calculation of cluster properties using the measured magnitudes and sharpness values from HSTphot.
\placefigure{fig_sampledv}
\placefigure{fig_samplesize}

Combining these two approaches, we can determine integrated magnitudes of the recovered artificial clusters using the analytical corrections and compare the clusters' true magnitudes to their measured magnitudes; this is shown in Figure \ref{fig_fakeclus}.  We found that the $V$ magnitudes are well-recovered (the mean error is $V_{recovered} - V_{input} = -0.01$, with rms scatter of 0.12 magnitudes), giving us confidence that this technique will produce reliable luminosity functions.
\placefigure{fig_fakeclus}

We then applied the analytical cluster measurement technique to our NGC 3627 objects, determining core radii and integrated magnitudes for the 553 candidate clusters.  We find that nine of the objects have recovered core radii greater than 1 pixel; these are eliminated, as we do not trust our stellar PSF-fitting photometry for objects larger than one pixel.  This cutoff also eliminates any small associations from the sample.  We also eliminated two objects that, from visual inspection, were not clusters (the nucleus of a background galaxy and part of a diffraction spike of a bright star).  No other obvious false detections were present.  Because of the low contamination level measured ($1-2$ false detections), we will henceforth refer to these 528 objects as ``clusters'' rather than ``candidate clusters''.

As a sanity check, we calculate the contribution of these clusters to the total light of NGC 3627.  We measure the combined light of the clusters to have magnitudes of $V_{clus} = 14.75$ and $I_{clus} = 14.03$.  In comparison, the integrated light in the WFPC2 field is $V_T = 10.38$ and $I_T = 9.43$, implying that the fractional contribution of the clusters is 1.8\% in $V$ and 1.4\% in $I$.  In order to compare with the results of Larsen \& Richtler (2000), we must count only the clusters with $M_V \le -9$, which account for 1.3\% of the $V$ luminosity and 1.0\% of the $I$ luminosity.  While this value is higher than those of the Larsen \& Richtler (2000) sample, we note that NGC 3627 is an interacting system and thus should be compared with the literature values they present for interacting systems.  Comparing with these systems, we find the cluster light fraction we find in NGC 3627 to fall comfortably within the range of values they give ($V$ fractions from 0.11\% for NGC 3921 to 15\% for NGC 3256).

\subsection{Cluster Population}

The properties of our detected clusters are shown in Figures \ref{fig_cluscmd} and \ref{fig_clussize}.  It is clear from Figure \ref{fig_cluscmd} that unambiguous interpretation of these objects in terms of determining age and initial mass (even if there is no differential reddening) is not possible, as $(V-I)$ is not a monotonic function of cluster age.  Rather, a cluster begins its life extremely blue, becomes redder when the first red supergiants form, moves slowly back to the blue, and eventually begins moving redward again.  Additionally, for young clusters smaller than $M_0 \approx 10^5 M_{\odot}$, the number of red supergiants at any given age is quite small, leading to a large intrinsic color dispersion for clusters of a single age.  Finally, we note that differential extinction is clearly present, as we see red clusters preferentially in and near the dust lanes.  For example, clusters a heavily-extincted region on WFC3 near the WFC2 border have a median $(V-I)$ color of 1.21, while those in a relatively unobscured region of WFC4 have a median color of $(V-I) = 0.49$.  If this color difference is caused solely by extinction, the reddening difference is $\Delta E(V-I) = 0.72$, or $A_V = 1.75$.  (The Cepheids presumably showed a smaller amount of differential extinction because they are much less luminous, and thus we are more biased against detecting highly-reddened Cepheids than we are against detecting highly-reddened clusters.)  Although one generally shows a cluster luminosity function, this is also difficult to interpret, as the rapid evolutionary dimming of a cluster makes it difficult to interpret the luminosity function in terms of cluster masses.  Between ages of 10 and 40 Myr, for example, the cluster dims by over a factor of 10 (the exact number depending, of course, on the assumed IMF).  Because of these difficulties in interpretation, the primary focus of our survey will be the measurement statistical properties of cluster populations of many systems, rather than an attempt to measure an accurate star formation history or cluster mass function of any one system.
\placefigure{fig_cluscmd}
\placefigure{fig_clussize}

The simplest population to isolate and study is the set of 13 clusters falling beyond the visual extent of the galaxy, which we define to be WFC4 objects with $y > 450$.  The positions and characteristics of these objects are shown in Table \ref{tab_halo}.  Interestingly, despite suffering little or no internal reddening in NGC 3627, these clusters have a median color of $(V-I) = 0.96$, much \textit{redder} than that of the other NGC 3627 clusters (0.69).  We thus conclude that these are predominantly old globular clusters, a conclusion strengthened by their having mean absolute magnitude ($\mean{M_V} = -8.1$; rms scatter of 1.0 magnitudes) comparable to that of Galactic globular clusters \citep{djo93}.  Very faint or small clusters would not have met our detection criteria.
\placetable{tab_halo}

Without a statistical CMD-synthesis approach (the subject of Paper II), the difficulties mentioned above, as well as incompleteness as a function of $r_c$ in addition to the usual $V$ and $I$, mean that we can only make a semi-quantitative study of Figures \ref{fig_cluscmd} and \ref{fig_clussize}.  We first note that there is a very strong correlation between positions of the clusters (Figures \ref{fig_chart0}-\ref{fig_chart3}) and the spiral structure of NGC 3627.  The spiral arm passing through the PC and WFC4, for example, contains 171 (32\%) of the detected clusters in just $\sim 10$\% of the WFPC2 field of view.  Adding a ring in WFC2 just outside the bulge and an inner spiral arm on the WFC2/WFC3 border, we find that a total of 80\% of the clusters are contained in 30\% of the image.  The concentration is highest for blue ($V-I < 0.5$) clusters, of which 92\% fall in these three regions.  The color ranges $0.5 < V-I \leq 0.75$, $0.75 < V-I \leq 1.0$, and $V-I > 1.0$ all have 75\% to 77\% of the clusters in the three regions.  The simplest conclusion from these data is that the clusters with $(V-I) \leq 0.5$ are a young population with minimal reddening, while the other clusters are a mixture of $\sim 75$\% reddened young clusters and $\sim 25$\% older clusters.
\placefigure{chart0}
\placefigure{chart1}
\placefigure{chart2}
\placefigure{chart3}

Figure \ref{fig_lfall} shows the raw luminosity functions for all clusters recovered; Figure \ref{fig_lfblue} shows only the 126 clusters with $(V-I) \le 0.5$.  Because of the shallower photometry limit in WFC2 we have split both diagrams into WFC2 and the other chips.  Using the deeper PC+WFC3/4 luminosity functions down to $M_V = -8$, we fit the luminosity functions to
\begin{equation}
\log N = ( 0.57 \pm 0.05 ) M_V + const
\end{equation}
for all clusters, and
\begin{equation}
\log N = ( 0.49 \pm 0.09 ) M_V + const
\end{equation}
for blue clusters only.  These fits correspond to luminosity function slopes of $d \log N / d \log L = -1.43 \pm 0.12$ for all clusters and $-1.22 \pm 0.23$ for blue clusters.  Although these values are more shallow than those found in NGC 4038/9 by \citet{whi99}, they are similar to the \citet{whi99} slopes before completeness corrections.  Completeness corrections, however, depend significantly on the assumed distribution of cluster sizes.  Using large ($0.75$ pc $< r_c < 2$ pc) clusters, we are essentially 100\% complete below $M_V = -8$, thus leaving the LF slopes unchanged.  Using small ($0.25$ pc $< r_c < 1$ pc) clusters, we are only $\sim 50\%$ complete at $M_V = -8$, and thus the measured LF slope becomes steeper by 0.25.  The actual value is likely somewhere in between; we will adopt a preliminary LF slope of $d \log N / d \log L = -1.53 \pm 0.15$ for this data set.  A detailed determination of the cluster initial mass function employing population synthesis techniques will be the subject of Paper II.
\placefigure{fig_lfall}
\placefigure{fig_lfblue}

\section{Summary}

The question of modes of cluster formation is presently an open one.  While massive young clusters that are potentially the progenitors of globular clusters have been observed only in galaxies with very high star formation rates (interacting and starburst systems), it is unclear whether this is caused by a different mode of star formation than that observed in our Galaxy, or whether it is simply an increase in cluster formation rates of clusters of all masses.  In order to answer this question, we have begun a survey of nearby galaxies in which we will study the cluster properties of each, as well as trends that appear as functions of galaxy properties.

Because robust photometric techniques for slightly-extended objects (core radius smaller than the PSF) are not yet established, we have detailed the difficulties and methods that we have developed for this task.  Although our techniques are optimized for the specific task of measuring slightly-extended clusters on top of a rapidly-varying background, we believe the principles used here can be applied to other similar problems.  Specifically, our sky measurement algorithm can be used for any photometry (stellar or extended-object) on a rapidly-varying background, while our use of sharpness to measure the image shape can be used to PSF determination as well as object size determination.

We have applied these techniques to NGC 3627, for which deep observations (34 orbits in F555W and F814W) are available in the HST archive.  We find that the stellar PSF-fitting package HSTphot works extremely well for our task, both in its ability to accurately photometer previously-detected Cepheids and in its usefulness in discriminating and photometering clusters.

For the Cepheids, we recover 28 Cepheids found in previous studies \citep{sah99,gib00}.  Measuring their periods and mean magnitudes, we find a distance consistent with those measured by those authors, but inconsistent with the final value used by \citet{fre01}.

We have located 528 clusters in NGC 3627, which account for $\sim 1-2$\% of the galaxy's total light in $V$ and $I$.  We find that the blue cluster population traces NGC 3627's spiral pattern extremely closely, while redder clusters less so.  This is likely because of a population of halo globular clusters, seen clearly beyond the visual extent of the disk but likely appearing throughout the image as foreground objects.  Red clusters are also preferentially seen in and near dust lanes; we interpret these as reddened young clusters.  We measure a preliminary cluster luminosity function of $d \log N / d \log L = -1.53 \pm 0.15$; a definitive result will come from the statistical analysis that will be presented in Paper II.

\acknowledgments

Support for this work was provided by NASA through grant number AR-09196.01 from the Space Telescope Science Institute.  All of the data presented in this paper were obtained from the Multimission Archive at the Space Telescope Science Institute (MAST). STScI is operated by the Association of Universities for Research in Astronomy, Inc., under NASA contract NAS5-26555.

\clearpage
\begin{figure}
%\plotone{figs/image.eps}
\caption{NGC 3627 deep image, created by combining all $F555W$ images taken at the second pointing (a total of 14 images).}
\label{fig_image}
\end{figure}

\begin{figure}
%\plotone{figs/psf0.eps}
\caption{HSTphot sharpness values vs. $V$ magnitude for all recovered objects in the planetary camera of one deep $F555W$ image, using library PSFs.  Cepheids found by \citet{sah99} and \citet{gib00} are marked with asterisks, and have a median sharpness value of $-0.10$.}
\label{fig_psf0}
\end{figure}

\begin{figure}
%\plotone{figs/psfngc121.eps}
\caption{HSTphot sharpness values vs. $V$ magnitude for all recovered objects in the planetary camera of an image centered on the SMC globular cluster NGC 121.}
\label{fig_psfngc121}
\end{figure}

\begin{figure}
%\plotone{figs/psff.eps}
\caption{HSTphot sharpness values vs. $V$ magnitude for all recovered objects in the planetary camera of one deep $F555W$ image, using adjusted PSFs.  Cepheids found by \citet{sah99} and \citet{gib00} are marked with asterisks, and have a median sharpness value of $-0.05$.}
\label{fig_psff}
\end{figure}

\begin{figure}
%\plotone{figs/psffake_envelope.eps}
\caption{HSTphot sharpness values vs. $V$ magnitude for all recovered artificial stars placed on the NGC 3627 images.  The plotted envelope is the $2.5 \sigma$ limit.  The relatively-symmetric nature of these plots indicates the blending, while non-zero, is not a significant factor.}
\label{fig_psffake_envelope}
\end{figure}

\begin{figure}
%\plotone{figs/psffake_round.eps}
\caption{HSTphot roundness values vs. $V$ magnitude for all recovered artificial stars placed on the NGC 3627 images.  The plotted envelope is the $2 \sigma$ limit.}
\label{fig_psffake_round}
\end{figure}

\begin{figure}
%\plotone{figs/psf_envelope.eps}
\caption{HSTphot sharpness values vs. $V$ magnitude for all detected objects in NGC 3627.  The plotted $2.5 \sigma$ envelope is taken from Figure \ref{fig_psffake_envelope}, expanded by $\pm 0.05$; the magnitude cutoff is intended to reduce contamination.}
\label{fig_psf_envelope}
\end{figure}

\begin{figure}
%\plotone{figs/psf_round.eps}
\caption{HSTphot roundness values vs. $V$ magnitude for all detected objects in NGC 3627.  The plotted $2 \sigma$ envelope is taken from Figure \ref{fig_psffake_round}.}
\label{fig_psf_round}
\end{figure}

\begin{figure}
%\plotone{figs/psffclus_envelope.eps}
\caption{HSTphot sharpness values vs. $V$ magnitude for all recovered artificial clusters with input core radii of 2.5 pc.  The $2.5 \sigma$ stellar envelope is taken from Figure \ref{fig_psffake_envelope}.}
\label{fig_psffclus_envelope}
\end{figure}

\begin{figure}
%\plotone{figs/psffclus_round.eps}
\caption{HSTphot roundness values vs. $V$ magnitude for all recovered artificial clusters with input core radii of 2.5 pc.  The $2 \sigma$ stellar envelope is taken from Figure \ref{fig_psffake_round}.}
\label{fig_psffclus_round}
\end{figure}

\clearpage
\begin{figure}
%\plotone{figs/complete_pc.eps}
\caption{Completeness fraction contours in the PC, as a function of input $V$ magnitude and core radius (in pc).  Several Galactic globular clusters are shown for comparison, as well as the young LMC clusters NGC 1818 and NGC 2157.}
\label{fig_complete_pc}
\end{figure}

\begin{figure}
%\plotone{figs/fakeclus.eps}
\caption{Input minus output $V$ magnitudes for recovered artificial clusters of core radii $0.5 \hbox{pc} < r_c < 5 \hbox{pc}$.  The average difference is $\Delta V = -0.01$, with a standard deviation of 0.12 magnitudes.}
\label{fig_fakeclus}
\end{figure}

\begin{figure}
%\plotone{figs/sampledv.eps}
\caption{$V$ corrections (true $V$ minus measured $V$) determined by our analytical corrections, as a function of measured $V$ magnitude and sharpness, for a typical background level in the PC.  The four lines correspond to sharpness values of -0.1 (the bottom), -0.2, -0.3, and -0.4 (the top).}
\label{fig_sampledv}
\end{figure}

\begin{figure}
%\plotone{figs/samplesize.eps}
\caption{Core radii determined by our analytical corrections, as a function of measured $V$ magnitude and sharpness, for a typical background level in the PC and the measured distance to NGC 3627.  The four lines correspond to sharpness values of -0.1 (the bottom), -0.2, -0.3, and -0.4 (the top).}
\label{fig_samplesize}
\end{figure}

\begin{figure}
%\plotone{figs/cluscmd.eps}
\caption{$(V-I, V)$ CMD of 528 clusters in NGC 3627.  The upper solid line corresponds to the evolutionary track of a cluster with an initial mass of $10^5 M_{\odot}$, at the distance ($\dist = 30.03$) and extinction ($A_V = 0.67$) measured for NGC 3627.  The lower line is the trrack for a cluster with an initial mass of $10^3 M_{\odot}$.}
\label{fig_cluscmd}
\end{figure}

\begin{figure}
%\plotone{figs/clussize.eps}
\caption{Core radii vs. $V$ magnitudes of 528 clusters in NGC 3627.}
\label{fig_clussize}
\end{figure}

\begin{figure}
%\plotone{figs/chart0.eps}
\caption{Unsharp masked F555W PC image.  Clusters are circled.}
\label{fig_chart0}
\end{figure}

\begin{figure}
%\plotone{figs/chart1.eps}
\caption{Same as Figure \ref{fig_chart0}, for WFC2.  The texture is the result of unsharp masking this very crowded field.}
\label{fig_chart1}
\end{figure}

\begin{figure}
%\plotone{figs/chart2.eps}
\caption{Same as Figure \ref{fig_chart0}, for WFC3.}
\label{fig_chart2}
\end{figure}

\begin{figure}
%\plotone{figs/chart3.eps}
\caption{Same as Figure \ref{fig_chart0}, for WFC4.}
\label{fig_chart3}
\end{figure}

\begin{figure}
%\plotone{figs/lfall.eps}
\caption{Raw luminosity functions (uncorrected for incompleteness) for all clusters found in NGC 3627.  The top panel shows WFC2 detections; the bottom panel shows detections in other chips.}
\label{fig_lfall}
\end{figure}

\clearpage
\begin{figure}
%\plotone{figs/lfblue.eps}
\caption{Raw luminosity functions (uncorrected for incompleteness) for blue ($V-I < 0.5$) clusters found in NGC 3627.  The top panel shows WFC2 detections; the bottom panel shows detections in other chips.}
\label{fig_lfblue}
\end{figure}

\clearpage
\begin{deluxetable}{llrrrr}
\tablecaption{Mean photometry values after the first iteration, as calculated using Cepheids found by previous studies. \label{tab_psf0}}
\tablewidth{0pt}
\tablehead{
\colhead{Image} &
\colhead{Chip} &
\colhead{$\mean{counts}$ ($e^{-}$)} &
\colhead{$\mean{bg}$ ($e^{-}$)} &
\colhead{$\mean{\hbox{sharp}}$} &
\colhead{$\delta FWHM$ (pixels)}}
\startdata
F555W-1 &  PC1 &  $3915$ &  $1957$ & $-0.103$ & $+0.516$ \\
F555W-1 & WFC2 & $13000$ & $12129$ & $-0.090$ & $+0.245$ \\
F555W-1 & WFC3 &  $7871$ &  $7903$ & $-0.083$ & $+0.227$ \\
F555W-1 & WFC4 &  $9028$ &  $9804$ & $-0.068$ & $+0.192$ \\
F555W-2 &  PC1 &  $5855$ &  $2383$ & $-0.079$ & $+0.391$ \\
F555W-2 & WFC2 & $18435$ & $15413$ & $-0.099$ & $+0.266$ \\
F555W-2 & WFC3 & $12402$ &  $9257$ & $-0.038$ & $+0.117$ \\
F555W-2 & WFC4 & $10940$ & $11773$ & $-0.102$ & $+0.278$ \\
F814W-1 &  PC1 &  $2488$ &   $978$ & $-0.114$ & $+0.618$ \\
F814W-1 & WFC2 &  $6548$ &  $5555$ & $-0.087$ & $+0.281$ \\
F814W-1 & WFC3 &  $4236$ &  $3737$ & $-0.073$ & $+0.248$ \\
F814W-1 & WFC4 &  $5329$ &  $3937$ & $-0.057$ & $+0.206$ \\
F814W-2 &  PC1 &  $3739$ &  $1320$ & $-0.104$ & $+0.545$ \\
F814W-2 & WFC2 & $10529$ &  $7448$ & $-0.080$ & $+0.258$ \\
F814W-2 & WFC3 &  $7163$ &  $4739$ & $-0.101$ & $+0.322$ \\
F814W-2 & WFC4 &  $6197$ &  $5127$ & $-0.008$ & $+0.077$ \\
\enddata
\end{deluxetable}

\clearpage
\begin{deluxetable}{llrrrr}
\tablecaption{Cepheids recovered in our photometry. \label{tab_cepheid}}
\tablewidth{0pt}
\tablehead{
\colhead{ID\tablenotemark{a}} &
\colhead{ID\tablenotemark{b}} &
\colhead{$V$} &
\colhead{$I$} &
\colhead{P(d)} &
\colhead{Q}\tablenotemark{c}}
\startdata
C1-V7  & --- & $25.21 \pm 0.06$ & $23.99 \pm 0.09$ & $42.9 \pm 4$ & 4 \\
C1-V6  & --- & $25.38 \pm 0.11$ & $24.14 \pm 0.12$ & $37.3 \pm 4$ & 3 \\
------ & C15 & $24.54 \pm 0.09$ & $23.49 \pm 0.11$ & $35.8 \pm 4$ & 3 \\
C2-V34 & C23 & $24.64 \pm 0.10$ & $23.42 \pm 0.12$ & $45.4 \pm 5$ & 4 \\
C2-V4  & C04 & $24.64 \pm 0.09$ & $23.57 \pm 0.11$ & $42.5 \pm 3$ & 4 \\
C2-V33 & --- & $24.64 \pm 0.11$ & $23.67 \pm 0.13$ & $30.2 \pm 3$ & 4 \\
------ & C20 & $24.67 \pm 0.12$ & $23.57 \pm 0.13$ & $34.2 \pm 5$ & 3 \\
C2-V8  & C10 & $24.90 \pm 0.13$ & $23.98 \pm 0.14$ & $41.0 \pm 6$ & 3 \\
C2-V31 & C11 & $25.32 \pm 0.12$ & $23.76 \pm 0.13$ & $33.2 \pm 4$ & 2 \\
C2-V26 & C05 & $25.23 \pm 0.12$ & $24.15 \pm 0.13$ & $28.3 \pm 3$ & 3 \\
C2-V12 & --- & $25.23 \pm 0.11$ & $24.27 \pm 0.13$ & $25.9 \pm 3$ & 4 \\
C2-V21 & C19 & $25.18 \pm 0.08$ & $24.06 \pm 0.12$ & $24.0 \pm 2$ & 4 \\
C2-V9  & --- & $25.28 \pm 0.22$ & $23.76 \pm 0.23$ & $21.5 \pm 4$ & 2 \\
C2-V35 & --- & $25.18 \pm 0.33$ & $24.39 \pm 0.35$ & $28.6 \pm 3$ & 2 \\
C2-V32 & --- & $25.61 \pm 0.16$ & $24.37 \pm 0.20$ & $23.1 \pm 4$ & 2 \\
C2-V15 & --- & $26.01 \pm 0.20$ & $24.98 \pm 0.22$ & $27.1 \pm 3$ & 3 \\
C3-V10 & C24 & $24.37 \pm 0.12$ & $23.40 \pm 0.13$ & $40.5 \pm 5$ & 4 \\
C3-V3  & C28 & $25.31 \pm 0.17$ & $24.15 \pm 0.18$ & $29.1 \pm 3$ & 4 \\
C3-V6  & --- & $25.53 \pm 0.12$ & $24.47 \pm 0.14$ & $21.9 \pm 2$ & 4 \\
C3-V17 & C26 & $25.54 \pm 0.11$ & $24.51 \pm 0.13$ & $25.8 \pm 3$ & 3 \\
C3-V4  & --- & $25.54 \pm 0.11$ & $24.50 \pm 0.13$ & $26.8 \pm 3$ & 4 \\
------ & C32 & $25.60 \pm 0.11$ & $24.43 \pm 0.13$ & $29.2 \pm 3$ & 3 \\
C3-V12 & C30 & $26.02 \pm 0.20$ & $24.74 \pm 0.21$ & $24.0 \pm 2$ & 4 \\
------ & C36 & $23.85 \pm 0.05$ & $22.76 \pm 0.07$ & $51.4 \pm 5$ & 2 \\
C4-V9  & C34 & $25.35 \pm 0.12$ & $24.23 \pm 0.14$ & $24.5 \pm 2$ & 4 \\
C4-V13 & --- & $25.48 \pm 0.21$ & $24.13 \pm 0.23$ & $26.4 \pm 3$ & 4 \\
C4-V4  & --- & $25.35 \pm 0.11$ & $24.47 \pm 0.13$ & $25.5 \pm 3$ & 3 \\
C4-V10 & C33 & $25.46 \pm 0.18$ & $24.45 \pm 0.20$ & $24.3 \pm 3$ & 3 \\
\enddata
\tablenotetext{a}{\citet{sah99}}
\tablenotetext{b}{\citet{gib00}}
\tablenotetext{c}{Light curve quality, on a scale of $0-4$.  Only 2 and higher are shown.}
\end{deluxetable}

\clearpage
\begin{deluxetable}{rrrrrr}
\tablecaption{Halo Clusters. \label{tab_halo}}
\tablewidth{0pt}
\tablehead{
\colhead{chip} &
\colhead{$x$} &
\colhead{$y$} &
\colhead{$r_c$ (pc)} &
\colhead{$M_V$\tablenotemark{a}} &
\colhead{$(V-I)_0$\tablenotemark{a}}}
\startdata
WFC4 &  56.78 & 777.82 & $2.53 \pm 0.08$ & $-7.00 \pm 0.02$ & $1.02 \pm 0.04$ \\
WFC4 &  86.17 & 607.53 & $1.72 \pm 0.07$ & $-6.70 \pm 0.02$ & $0.79 \pm 0.03$ \\
WFC4 & 121.05 & 772.64 & $1.70 \pm 0.07$ & $-6.58 \pm 0.02$ & $0.90 \pm 0.03$ \\
WFC4 & 150.78 & 473.85 & $4.19 \pm 0.09$ & $-8.26 \pm 0.02$ & $1.10 \pm 0.02$ \\
WFC4 & 152.78 & 731.71 & $1.89 \pm 0.04$ & $-7.82 \pm 0.01$ & $0.74 \pm 0.01$ \\
WFC4 & 172.78 & 532.45 & $2.09 \pm 0.06$ & $-7.48 \pm 0.01$ & $0.57 \pm 0.02$ \\
WFC4 & 383.22 & 551.29 & $2.37 \pm 0.05$ & $-7.77 \pm 0.01$ & $1.01 \pm 0.02$ \\
WFC4 & 393.84 & 771.25 & $2.06 \pm 0.07$ & $-6.87 \pm 0.02$ & $1.00 \pm 0.03$ \\
WFC4 & 420.08 & 581.83 & $2.14 \pm 0.05$ & $-7.82 \pm 0.01$ & $0.69 \pm 0.02$ \\
WFC4 & 428.31 & 567.79 & $2.84 \pm 0.06$ & $-7.92 \pm 0.01$ & $0.65 \pm 0.02$ \\
WFC4 & 437.08 & 499.18 & $3.74 \pm 0.13$ & $-7.38 \pm 0.02$ & $1.02 \pm 0.04$ \\
WFC4 & 532.17 & 569.17 & $4.55 \pm 0.13$ & $-7.85 \pm 0.02$ & $1.16 \pm 0.03$ \\
WFC4 & 735.99 & 511.08 & $2.11 \pm 0.05$ & $-7.59 \pm 0.01$ & $0.71 \pm 0.03$ \\
\enddata
\tablenotetext{a}{Corrected only for foreground extinction of $A_V = 0.11$ \citep{sch98}}
\end{deluxetable}

\end{document}